%% file: conference_101719.tex
\documentclass[conference]{IEEEtran}
\IEEEoverridecommandlockouts
\usepackage{amsmath,amssymb,amsfonts}
\usepackage{algorithmic}
\usepackage{graphicx}
\usepackage{textcomp}
\usepackage[HTML]{xcolor}
\usepackage[style=ieee]{biblatex}
\def\BibTeX{{\rm B\kern-.05em{\sc i\kern-.025em b}\kern-.08em
    T\kern-.1667em\lower.7ex\hbox{E}\kern-.125emX}}
\usepackage{tikz}
\usepackage{algorithm2e}
\usepackage{multirow}

\newcommand{\R}{\mathbb{R}}
\newcommand{\C}{\mathbb{C}}

\renewcommand{\vec}[1]{\mathbf{#1}}
\newcommand{\grad}{\nabla}

\begin{document}

\title{Efficient Domain Decomposition for the Helmholtz Equation on GPUs.
\thanks{This material is based upon work supported by the National Science Foundation under Grant Numbers DMS-2345225 and DMS-2436319 and Virginia Tech.}
\thanks{This work has been submitted to IEEE for possible publication. Copyright may be transferred without notice, after which this version may no longer be accessible.}
}

\author{\IEEEauthorblockN{1\textsuperscript{st} Amit Rotem}
\IEEEauthorblockA{\textit{Department of Mathematics} \\
\textit{Virginia Tech}\\
Blacksburg, USA \\
arotem@vt.edu}

}

\maketitle

\begin{abstract}
The Helmholtz equation governs wave propagation in acoustics, electromagnetics, and seismology, but its indefinite nature makes it difficult to solve with iterative methods. Domain decomposition methods are a natural fit for massively parallel architectures, yet mapping efficient Helmholtz solvers onto modern GPUs remains a challenge. We address both with two key contributions: (1) a block-level domain decomposition scheme, in which each subdomain is assigned to a single thread block and all solves run concurrently in a single kernel launch, and (2) WaveHoltz as the subdomain solver. WaveHoltz is a fixed-point iteration that is uniquely well-suited to the GPU execution model due to its minimal memory footprint and no reduction operations. Together, these eliminate device-level synchronizations and replace global memory traffic with shared memory and register-level operations, keeping subdomain data largely resident in L1 and L2 cache.
We explore two threading strategies: one degree of freedom per thread for small subdomains, and multiple degrees of freedom per thread for larger ones.
Benchmarks of our CUDA based implementation on a NVIDIA A100 show that WaveHoltz achieves 2x–25x speedup over MINRES, with the advantage growing with subdomain size. Crucially, evaluating the subdomain solver in single rather than double precision yields an additional 2x–10x speedup--a benefit largely unattainable by MINRES due to loss of Krylov vector orthogonality under reduced precision.
\end{abstract}


\begin{figure}
    \centering
    \includegraphics[width=\linewidth]{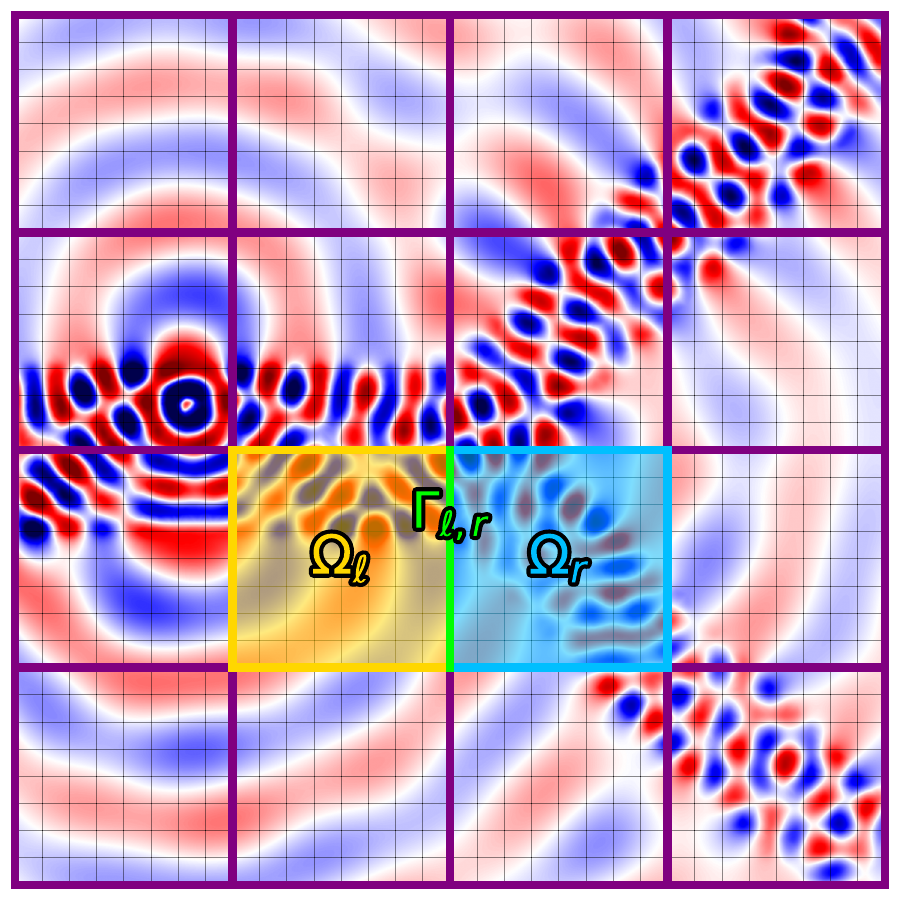}
    \caption{A partition of a rectangular domain $\Omega\subset\R^2$ into several subdomains. The elements are outlined in black, and the subdomains are outlined in purple. Two representative subdomains $\Omega_{\ell}$ and $\Omega_{r}$ and their common interface $\Gamma_{\ell,r}$ are highlighted. The solution visualized is the wave produced by an off-center point source in a Y-shaped wave guide with a $3\times$ contrast in wavespeed relative to the surrounding media. The solution is computed with degree 3 basis functions using the method described here.}
    \label{fig:subdomains}
\end{figure}

\section{Introduction}
Due to its indefinite nature, the Helmholtz equation is difficult to solve with iterative methods \cite{Helmholtz-Difficult-Ernst-Gander, Advancements-Iterative-Helmholtz-Erlangga}. Krylov space methods, such as MINRES/GMRES, converge very slowly without a good preconditioner. Other iterative methods such as an off-the-shelf multigrid do not converge for this problem and must be nontrivially specialized to the Helmholtz equation. Algorithms based on factorization, such as multifrontal methods, or ILU and sweeping preconditioners can be very effective, but the inherent sequentialism and superlinear costs of these methods makes them less suitable for massively parallel GPU computing.
Time domain methods such as controllability methods \cite{grote2019controllability, GROTE2020112846} and WaveHoltz \cite{WaveHoltz, ROTEM2026114882, Appel2025AnOO} are also effective.

The method we present here is a non-overlapping domain decomposition method.
Domain decomposition methods (DDMs) are reliably efficient for solving the Helmholtz equation and naturally parallelize to many cores. The Helmholtz equation is indefinite and DDMs that solve it differ slightly from related methods for positive definite problems, mainly in the choice of transmission conditions between subdomains (discussed further in Section \ref{sec:DD}.) The first method extending DDMs to the Helmholtz equation was presented in \cite{Domain-Decomp-Depres}. Recent advancements towards making DDMs more efficient typically target the convergence rate of the iteration. One class of methods aim to improve the transmission condition, for example, by incorporating perfectly matched layers \cite{TAUS2020109706, CSTOLK2013240, ROYER2022115006, doi:10.1137/16M109781X, Nataf-ABC} or approximately absorbing boundary conditions (ABCs) \cite{HAGSTROM1988299, Nataf-ABC, OptimizedSchwarz, CHEVALIER1998769, BOUBENDIR2012262}. The latter of which can be further improved by optimizing the coefficients of the ABC for the given problem \cite{OptimizedSchwarz}. Another class of methods improve the convergence rate of the iteration by a two-level coarse-space correction also known as deflation. The deflation vectors are computed either as part of the Krylov iteration \cite{gmres-dr, gcro-dr, block-iterative-recycling} or by choosing a suitable ansatz space such as planewaves or eigenvectors of the transmission operator, see for example \cite{CONEN201483, AUBRY2012155, KIMN20071507}. The convergence rate of DDMs scales adversely with the number of subdomains, but coarse-space methods effectively overcome this difficulty \cite{block-iterative-recycling, CONEN201483}. The best performing DDMs typically employ both improved transmission conditions and coarse-space correction.

In the context of high performance computing, domain decomposition naturally scale to many core environments. A detailed account is given in \cite{scalable-ddm}. Typically, one subdomain is assigned to one node and communication takes place between nodes that share an interface. Compared to a Krylov method (without a preconditioner) for the same problem distributed the same way, DDMs tend to be very efficient not only because they take far fewer iterations but also because the work between communication stages is much greater. To make the latter statement true, the work per CPU must be large. For example, on the order of one hundred thousand DOFs per core were used in the experiments conducted in \cite{scalable-ddm}.

On GPUs, this requirement is pronounced. When used effectively a single GPU can process massive amounts of floating point operations compared to a single CPU node and therefore must be saturated with much more work to outweigh communication costs. To that end, there are two immediately obvious strategies to make domain decomposition strategies efficient on GPUs: (A) larger subdomains, or (B) multiple subdomains on one GPU. Strategy (A) is merited by the fact that DDMs with a few large subdomains will converge faster than those with many small subdomains. This comes at the cost of solving a larger Helmholtz equation on each subdomain which, as already discussed, is difficult and the cost may not scale optimally with the problem size.
Here we consider strategy (B) which in turn can be used as an efficient inner solver for strategy (A).

To make strategy (B) effective, we propose taking moderately sized subdomains consisting of several hundred to a few thousand degrees of freedom (DOFs) and solving each subdomain problem using a thread block consisting of enough threads so that each thread is associated with a small fixed number of DOFs.

We present two variants of this strategy. First, we assign one thread to one DOF and the subdomains are taken small enough (fewer than 1024 DOFs) to store the entire subdomain in shared memory. The resulting kernel has a high arithmetic intensity and nearly all data movement occurs between registers and L1 cache. In the second variant, we have one thread per $T$ DOFs (2 or 4, say), and use a smaller shared memory buffer together with a large global buffer. This variant can solve problems on larger subdomains (fewer than $1024\times T$ DOFs) at the cost of increased cache and register pressure.

The alternative to these strategies is to distribute each subdomain over several thread blocks. For large subdomains, doing so may be practically unavoidable but comes at the cost of device level synchronizations. 
For such a configuration, iterative methods are less optimal.
Solving the subdomain problem may require hundreds of kernel launches compared to just one with the strategy we propose. When split over many kernels, the arithmetic intensity decreases and data movement dominates.
If larger subdomains are ultimately necessary, then it may be more advantageous to employ different strategies such as direct factorization methods, or using our proposed method hierarchically, but we do not explore this here.

Our approach is especially effective in two dimensions where subdomains with a few hundred degrees of freedom are large relative to the global problem, so the global system converges quickly. In three dimensions, these subdomains are notably smaller relative to the global problem, so other global acceleration strategies (such as coarse-space correction) are needed to make the solver effective.

The paper is organized as follows. In Sections \ref{sec:math}--\ref{sec:FEM} we introduce the Helmholtz equation, the non-overlapping domain decomposition method, and its spectral element discretization. In Section \ref{sec:WH} we describe the WaveHoltz iteration for the subdomain solve, and in Section \ref{sec:CUDA} we combine these components into a GPU algorithm and propose two threading strategies. Finally, Section \ref{sec:benchmarks} benchmarks the proposed method n two and three dimensions and compares WaveHoltz with MINRES as subdomain solvers.

\section{Background}\label{sec:math}

Consider the Helmholtz equation in a bounded and simply connected domain $\Omega \subset\R^d$ ($d = 2,3$):
\begin{subequations} \label{eq:Helmholtz}
    \begin{align}
        -\Delta u - k^2 u &= f, & x &\in \Omega, \label{eq:helmholtz-pde} \\
        \partial_nu - i k u &= \lambda, & x &\in \partial\Omega. \label{eq:imepdance}
    \end{align}
\end{subequations}
Here $k = \omega / c(x)$ where $\omega$ is the time-harmonic frequency and $c(x) > 0$ is the wave speed. The function $f = f(x)$ is a compactly supported forcing term (typically real-valued) and $\lambda = \lambda(x)$ is complex-valued.
When $\lambda = 0$, the impedance boundary condition approximates the Sommerfeld radiation condition
\[ \lim_{r\to \infty} r^{\frac{d-1}{2}} \left(\partial_r u - i k u \right) = 0, \quad r = |x|. \]
This radiation condition enforces the intuitive idea that all waves propagate out from the source $f$ and that no waves are propagating from the far-field.
More accurate approximations of the radiation conditions such as perfectly matched layers, or high order absorbing boundary conditions (see \cite{Nataf-ABC, NotesOnPML, EM-PML}) can meaningfully improve the quality of the solution, but here we limit our consideration to the impedance condition \eqref{eq:imepdance}.
Nevertheless, it is important to note that such boundary conditions are particularly relevant to domain decomposition methods which discuss we further in Section \ref{sec:DD}.

\section{Domain Decomposition}\label{sec:DD}
In this section, we broadly introduce the parallel non-overlapping domain decomposition methods for the Helmholtz equation. Define $L$ non-overlapping subdomains $\Omega_{\ell}$ for $\ell=1,\dots,L$ such that $\bigcup_{\ell=1}^L \Omega_{\ell} = \Omega$ and $\Omega_{\ell} \cap \Omega_{r} = \Gamma_{\ell,r} = \Gamma_{r,\ell}$ where $\Gamma_{\ell,r}$ is a $d-1$ dimensional interface such as in Figure \ref{fig:subdomains}. 

We define $u_{\ell}$ as the restriction of $u$ to subdomain $\Omega_{\ell}$. The domain decomposition method is iterative, and at iteration $n+1$ we update the solution $u_{\ell}^{n+1}$ by solving a Helmholtz equation on $\Omega_{\ell}$. Namely,
\begin{align*}
    -\Delta u^{n+1}_{\ell} - k^2 u^{n+1}_{\ell} &= f, & & x\in\Omega_{\ell} \\
    \mathcal{B}_{\ell, r} u_{\ell}^{n+1} &= \mathcal{B}_{\ell, r} u_r^{n}, & & x\in\Gamma_{\ell,r}\setminus\partial\Omega, \\
    (\partial_{n} - i k) u_{\ell}^{n+1} &= 0, & & x\in\partial\Omega.
\end{align*}
The transmission condition in terms of the operators $\mathcal{B}_{\ell,r}$ on $\Gamma_{\ell,r}$ characterize the particular domain decomposition method and the choice of operator here critically affects the convergence rate of the iteration. In the classical parallel Schwarz method $\mathcal{B}_{\ell,r} = \rm{Id}$, but this choice generally leads to a non-convergent iteration for the Helmholtz equation and the subdomain problems may not be well posed \cite{IntroDomainDecomp}. The theoretically optimal choice is to take $\mathcal{B}_{\ell,r} = \partial_n + DtN$ where $DtN$ is the Dirichlet-to-Neumann (also known as the Poincar\'{e}-Steklov) operator. For a partition of $\Omega$ into $L$ vertical slices, say, the iteration with the $DtN$ transmission condition converges in exactly $L$ iterations. However, the $DtN$ is a non-local operator and cannot be discretized efficiently. 
There are many ways to approximate the $DtN$ by local operators namely by ABCs just as discussed in Section \ref{sec:math} for the Sommerfeld radiation condition, see \cite{TAUS2020109706, ROYER2022115006, Nataf-ABC, OptimizedSchwarz}. A good choice of boundary conditions can dramatically improve the convergence rate \cite{doi:10.1137/16M109781X}. In this paper, we are concerned moreso with the efficient implementation of these methods on GPUs, so for simplicity we take the so called zero-th order transmission condition
\[
\mathcal{B}_{\ell,r} = \partial_n - i k.
\]

Enforcing the interface condition requires evaluating the normal derivatives of the solution at the boundary. This proves to be problematic for element based methods for which the derivatives are not necessarily continuous. To resolve this issue, we introduce the dual variables $\lambda_{\ell,r}$ for each pair of neighboring subdomains $(\Omega_{\ell}, \Omega_{r})$ such that on $\Gamma_{\ell,r}$
\[ \lambda_{\ell,r} = (\partial_{n} - i k) u_r, \quad \lambda_{r,\ell} = (\partial_{n} - i k) u_{\ell}. \]
We then solve the subdomain problems with the interface conditions:
\begin{align*}
    (\partial_{n} - i k) u_{\ell}^{n+1} &= \lambda_{\ell,r}^n, & & x\in\Gamma_{\ell,r}.
\end{align*}
It can be verified that from their definition (see \cite{IntroDomainDecomp}) that the iterates $\lambda^n_{\ell,r}$ satisfy:
\begin{align}\label{eq:lambda-iteration}
    \lambda_{\ell, r}^{n+1} &= -\lambda_{r,\ell}^n + 2 i k u_{r}^{n+1}\big|_{\Gamma_{\ell,r}}.
\end{align}
Note that rather than iterating on $u_{\ell}$ directly, we can equivalently iterate on $\lambda_{\ell,r}$ instead. This fixed point iteration can be accelerated with a Krylov space method and the resulting method is equivalent to preconditioning the global system \cite{IntroDomainDecomp}.

\section{Spectral Element Discretization}\label{sec:FEM}
We now present the weak formulation of \eqref{eq:Helmholtz}. We assume that the domain is partitioned into convex quadrilateral or hexahedral elements which each element belonging to only one subdomain. We assume quadrilateral or hexahedral elements to exploit the tensor product structure of the basis functions.
We seek ${u \in H^1(\Omega)}$ such that for all $\phi\in H^1(\Omega)$
\begin{equation}\label{eq:weak-helmholtz}
    (\grad u, \grad \phi) - (k^2 u, \phi) - \langle k u, \phi \rangle = (f, \phi) + \langle \lambda, \phi \rangle.
\end{equation}
Here,
\[ (a, b) = \int_\Omega a(x) \overline{b(x)} \, dx, \qquad \langle a, b \rangle = \int_{\partial\Omega} a(x) \overline{b(x)} \, ds. \]
After discretizations by the spectral element method \cite{Kopriva-ImplementingSpectralMethods} we are left with:
\begin{equation}\label{eq:discrete-helmholtz}
    (A - i\omega H - \omega^2 M) \vec{u} = \vec{b}.
\end{equation}
Here the matrices $A, M, H \in \C^{N \times N}$ and the vector $\vec{b}\in \C^N$ are defined as follows:
\begin{align*}
    A_{ij} &= ( \nabla \phi_j,\; \nabla \phi_i ), &
    M_{ij} &= ( c^{-2}\phi_j,\; \phi_i ), \\
    H_{ij} &= \langle c^{-1} \phi_j,\; \phi_i \rangle, &
    \vec{b}_j &= (f,\; \phi_j) + \langle \lambda ,\; \phi_j \rangle,
\end{align*}
where $\{\phi_i\}_{i=1}^N$ is the basis for the spectral element method.
It will prove useful in Section \ref{sec:WH} to approximate the integrals with the GLL quadrature rule on which the nodal basis is defined resulting in $M$ and $H$ being diagonal (mass-lumped).
We can also formulate \eqref{eq:discrete-helmholtz} as a real block system for the real and imaginary parts of $\vec{u}$:
\begin{equation}\label{eq:matrix-system}
    \begin{pmatrix}
        A - \omega^2 M & \omega H \\
        \omega H & \omega^2 M - A
    \end{pmatrix}
    \begin{pmatrix}
        \Re\{\vec{u}\} \\
        \Im\{\vec{u}\}
    \end{pmatrix}
    =
    \begin{pmatrix}
        \Re\{\vec{b}\} \\
        -\Im\{\vec{b}\}
    \end{pmatrix}.
\end{equation}
From the definitions of $A, M,$ and $H$ it is clear that the block system in \eqref{eq:matrix-system} is symmetric, nevertheless it is indefinite making it difficult for iterative methods to converge.
We also note that the matrices $A, M,$ and $H$ or the block system \eqref{eq:matrix-system} need not be explicitly formed. The linear system can be solved iteratively with a Krylov method with matrix-vector products computed by evaluating integrals on the fly via quadrature. For more details, we refer to \cite{matfree-fem-gpu} and \cite{ABDELFATTAH2021102841} wherein finite element operators are computed matrix-free on GPUs.

\section{Solving the Subdomain Problem: WaveHoltz}\label{sec:WH}
On each subdomain $\Omega_{\ell}$, \eqref{eq:Helmholtz} is discretized with the spectral element method as described in Section \ref{sec:FEM} which is equivalent to solving \eqref{eq:discrete-helmholtz} or \eqref{eq:matrix-system}. Among the fastest serial methods for solving this subdomain problem is to pre-compute a sparse factorization of \eqref{eq:discrete-helmholtz} associated with each subdomain and perform the sparse triangular solves at every iteration. This method works very well in message passing parallelism because the work per node is large relative to the communication cost \cite{scalable-ddm}. While there are many efforts to accelerate sparse triangular solves on GPUs, for example \cite{GHYSELS2022102897} and packages like CuDSS \cite{cudss}, the process is inherently sequential. To take advantage of the highly parallel nature of the GPU, we propose that the linear system should be solved iteratively (hopefully in a small number of iterations). In Section \ref{sec:CUDA} we propose solving each subdomain problem on a single thread block consisting of several hundred threads, so the iterative method we use should be highly parallelizable with a small memory footprint as the storage will occupy registers and shared memory which are highly limiting resources on the GPU. The Krylov space methods MINRES and BiCGSTAB meet our requirements \cite{Saad-IterativeMethods} and in Section \ref{sec:benchmarks} we show performance results using MINRES as it is typically more reliable than BiCGSTAB and takes advantage of the symmetry of the problem at roughly the same cost. Here, we proceed with the WaveHoltz iteration \cite{WaveHoltz} which has a smaller memory footprint than both MINRES or BiCGSTAB, requiring only one additional vector to carry out the iteration where MINRES requires five vectors. A second benefit of the WaveHoltz method is that it does not require any reduction operations (inner products) greatly reducing the number of thread synchronization steps.

The WaveHoltz iteration solves the Helmholtz equation by filtering the solution of the wave equation in time. Starting from \eqref{eq:discrete-helmholtz}, let $\vec{w} = \vec{u} e^{-i\omega t}$, then $\vec{w}$ solves:
\[
M \ddot{\vec{w}} + H \dot{\vec{w}} + A \vec{w} = \vec{b} e^{-i \omega t},
\]
with initial conditions $\vec{w}(0) = \vec{u}$ and $\dot{\vec{w}}(0) = -i\omega \vec{u}$. In fact, it is sufficient to solve for just the real part of $\vec{w}$, say $\vec{p} = \Re\{\vec{w}\}$, then
\begin{equation} \label{eq:diff-eq}
    M\ddot{\vec{p}} + H\dot{\vec{p}} + A \vec{p} = \Re\{\vec{b} e^{-i \omega t}\}.
\end{equation}
The WaveHoltz iteration solves for $\vec{u}$ by finding initial conditions to $\vec{p}$ so that $\vec{p}$ is periodic with frequency $\omega$. We start with an initial guess $\vec{u}$ and perform the fixed point iteration
\begin{equation}\label{eq:WaveHoltz}
    \vec{u}^{(m+1)} = \Pi \vec{u}^{(m)} = \frac{2}{N_t} \sum_{\nu=0}^{N_t-1} (\kappa_{\nu} \vec{p}^{\nu} - \tfrac{1}{i\omega}\kappa_{\nu+\frac{1}{2}} \dot{\vec{p}}^{\nu+\frac{1}{2}}).
\end{equation}
Here $N_t$ is the number of time steps, $\kappa_\nu = \cos\frac{2\pi \nu}{N_t} - \frac{1}{4}+\frac{1}{4}\tan^2\frac{\pi}{N_t}$ is the filter, and
$\vec{p}^\nu = \vec{p}(\frac{2\pi \nu}{N_t})$ where $\vec{p}(t)$ solves \eqref{eq:diff-eq} with initial condition $\vec{p}(0) = \Re\{\vec{u}^{(m)}\}$ and $\dot{\vec{p}}(0) = \Re\{-i\omega \vec{u}^{(m)}\}$. The filter $\kappa_\nu$ selects for the $\omega$-frequency component of $\vec{p}(t)$. The wave equation is evolved with the following modified leapfrog scheme. For the $j$-th component,
\begin{gather*}
    \dot{\vec{p}}^{\nu+\frac{1}{2}}_j = \alpha_j\dot{\vec{p}}^{\nu-\frac{1}{2}}_j + \beta_j \left[-(A\vec{p}^\nu)_j + \Re\{\vec{b}_j e^{-2\pi i \nu / N_t}\} \right], \\
    \vec{p}_j^{\nu+1} = \vec{p}^\nu_j + \sigma\, \dot{\vec{p}}^{\nu+\frac{1}{2}}_j.
\end{gather*}
Here $\alpha_j = \frac{M_{jj} - \theta H_{jj}}{M_{jj} + \theta H_{jj}}$ and $\beta_j = \frac{\sigma}{M_{jj} + \theta H_{jj}}$ where $\theta = \frac{\tan(\omega\Delta t / 2)}{\omega}$ and $\sigma = \frac{\sin(\omega \Delta t/2)}{\omega / 2}$. Observe that $\theta \approx \Delta t/2$ and $\sigma \approx \Delta t$ are chosen so that if $\vec{p}^0 = \Re\{\vec{u}\}$ and $\dot{\vec{p}}^{\frac{1}{2}} = \Re\{-i\omega \vec{u} e^{-i \omega \Delta t/ 2}\}$ then $\vec{p}^\nu = \Re\{\vec{u} e^{-i \nu \omega \Delta t}\}$ and $\dot{\vec{p}}^{\nu+\frac{1}{2}} = \Re\{-i\omega \vec{u} e^{-i (\nu + \frac{1}{2}) \omega \Delta t}\}$. That is, this scheme is designed to exactly evolve $e^{-i\omega t}$ for fixed $\omega$. This property, together with the time-filter $\kappa_\nu$ guarantee that the WaveHoltz iteration exactly converges to $\vec{u}$.
Note also that this scheme relies on $M$ and $H$ being diagonal so that the scheme is effectively explicit.
This scheme has a time-stepping restriction $\Delta t \leq \frac{2h}{p^2c_{\rm max}}$ where $h$ is the smallest characteristic element length, $p$ is the degree of the basis, and $c_{\rm max}$ is the fastest wavespeed in $\Omega$.

In \cite{ROTEM2026114882}, it was demonstrated that WaveHoltz can be interpreted as a preconditioned iteration for \eqref{eq:discrete-helmholtz}. For large problems the WaveHoltz iteration can be further accelerated with a Krylov space method and preconditioned with deflation making it robust and efficient solver on its own; see \cite{Appel2025AnOO} where also implicit time-stepping is used.
However, to reduce memory and syncronization cost, we do not accelerate the iteration in this work.
We find that very few (five to twenty) iterations are sufficient to converge to machine precision for the sizes of subdomains in our method. One might also consider solving \eqref{eq:discrete-helmholtz} on each subdomain to a low tolerance initially, then tightening the tolerance as \eqref{eq:lambda-iteration} approaches convergence to reduce total runtime. We do not investigate this strategy here, but report that significant progress towards convergence can be made with as few as two WaveHoltz iterations.

\section*{Summary of the DDM Cycle}
One step of the iteration \eqref{eq:lambda-iteration} on one subdomain is summarized in Algorithm \ref{alg:dd-subroutine}. It is clear from the pseudocode that to solve the subdomain problem, we need only store a complex vector $\vec{u}$ (or equivalently two real vectors) and two real vectors $\vec{p}$ and $\dot{\vec{p}}$ The operations on the vector $\vec{v}$ can be performed inplace on $\vec{u}$. Similarly, $A\vec{p}$ need not actually be stored, the result of the matrix-vector multiplication can be added directly to $\dot{\vec{p}}$. In comparison, an efficient MINRES implementation will require five complex vectors and two inner products per iteration. We could monitor the convergence of WaveHoltz by checking the residual of the fixed-point iteration, but it is worth noting that far fewer inner products are needed by WaveHoltz compared to MINRES because WaveHoltz requires far fewer iterations at the cost of more work per iteration.

\begin{algorithm}
    \caption{(DDM Cycle) Domain decomposition subroutine on subdomain $\Omega_{\ell}$.}
    \label{alg:dd-subroutine}
    \input{DDAlg}
\end{algorithm}

\section{GPU Algorithms}\label{sec:CUDA}
In this section, we summarize the design of the domain decomposition method for the GPU. The ultimate goal is to exploit as much parallelism as possible in the domain decomposition method. At each outer iteration, we apply the DDM cycle of Algorithm \ref{alg:dd-subroutine}. Since we can apply the DDM cycle to all subdomains concurrently, we identify this as the highest level parallelism we can exploit. Indeed this is the key feature that makes domain decomposition methods scalable \cite{IntroDomainDecomp, scalable-ddm}. By solving these linear systems with an iterative method as described in Section \ref{sec:WH} we expose much finer grained parallelism through the computation of vector operations and in the evaluation of SEM operators.

Synchronizations are ultimately unavoidable in the evaluation of the SEM operators, namely, the evaluation of $A\vec{p}$. There are several levels of synchronizations that can take place on a GPU; warp-level, block-level, and device-level synchronizations all come with their own costs. Warp-level synchronizations are essentially free by nature of how the GPU schedules instructions to threads. Block-level synchronizations are slower, but far slower are device-level synchronizations.

Looking at the data dependence in Algorithm \ref{alg:dd-subroutine}, we observe that synchronization must occur between WaveHoltz iterations, between time-stepping iterations, and during the computation of $A \vec{p}$. If the DDM cycle is implemented at the device level, that is, a single domain is treated by more threads than the maximum number of threads per thread block, then the number of independent kernel calls must scale like $O(N_t)$ per WaveHoltz iterations (for MINRES or a similar iterative solver, fewer kernel calls are needed per iteration, but many more iterations are needed).
The arithmetic intensity per time-step is low and each kernel call must read and write to $\vec{p}$ and $\dot{\vec{p}}$ which necessarily occupy global memory.
If each subdomain can be treated by a single thread block then this data movement is redundant. Instead we advocate that the entire DDM cycle be executed by a single thread block in a single kernel call. Then the number of reads and writes to global memory are substantially reduced and replaced with reads and writes to shared memory.
To emphasize this point, note that NVIDIA's A100 GPU used in our benchmarks in Section \ref{sec:benchmarks} has a latency of approximately 300 cycles for global memory, 200 cycles for L2, and 20-30 cycles for L1/Shared memory as reported in \cite{Abdelkhalik2022DemystifyingTN}.
Moreover since all compute operations are fused into one kernel the compute cycles dominate the runtime so the GPU is being effectively used.

To perform the DDM cycle on a thread block, we explore two threading strategies. First we assign one degree of freedom (DOF) to each thread. Second, we assign $T$ DOFs to each thread with $T$ small. In our examples we take $T=2$ and 4. We now describe these strategies.

\subsection{One DOF per Thread}\label{sec:T=1}
On every NVIDIA GPU since 2010 only 1024 threads can be grouped into one thread block implying that subdomains of at most $1024$ DOFs can be considered for our proposed strategy.
To evaluate $A \vec{p}$ matrix-free, it is more practical to describe the method in terms of elements rather than individual DOFs. To that end, assume a subdomain has $K$ elements of degree $P$ and assign to each thread block $K (P+1)^d$ threads. While $K(P+1)^d$ exceeds the number of DOFs, the redundancy in threads is worthwhile for computing $A\vec{p}$ matrix-free as it is by far the most expensive operation.

Note that the number of threads needed in three dimensions per element is substantially greater than in two dimensions limiting the number of elements that can be assigned to each subdomain. It is not necessary that each subdomain have exactly $K$ elements. We assume that all subdomains have at most $K$ elements and excess threads go unused\footnote{To avoid warp divergence, the DOF vector is extended by zeros and the excess threads accumulate zeros.}.

At the start of Algorithm \ref{alg:dd-subroutine}, all threads read the right hand side $f$ from global memory, and the threads associated with boundary DOFs additionally read the transmitted values $\lambda$. Each thread stores a single entry $\vec{b}_{j}$ assembled from $f$ and $\lambda$ independently in registers. The threads store single entries of $\vec{p}, \dot{\vec{p}},$ and $\vec{u}$ and their corresponding entry of $\alpha$ and $\beta$ in registers. Note that all steps in Algorithm \ref{alg:dd-subroutine} besides the matrix-vector product $A\vec{p}$ are entry-wise operations that can be carried out by the threads concurrently without any branching, synchronizations, or data movement outside of registers.

Computing $A \vec{p}$ requires synchronization and the sharing of memory between threads.
The shared memory consists of three buffers, the 1D differentiation matrix $D$ of size $(P+1)^2$, a buffer \texttt{pShared} of length $K (P+1)^d$ where the solution is stored, a buffer \texttt{gradShared} of length $d K (P+1)^d$ where the gradient is stored. Shared memory usage can be further reduced at the cost of one additional synchronization per evaluation of $A\vec{p}$ if \texttt{pShared} aliases \texttt{gradShared}.

When evaluating $A\vec{p}$, each thread maps its local entry of $\vec{p}$ to its corresponding spot in \texttt{pShared} and the threads synchronize. Each thread computes the gradient of $\vec{p}$ at one quadrature point, which consists of $d$ inner products with the corresponding rows of $D$. The threads write to \texttt{gradShared} then synchronize. Each thread accumulates the integral $(\grad p, \grad \phi)$ which consists of another $d $ inner products with rows of $D^T$. The result is accumulated in \texttt{pShared} atomically and the threads synchronize. The atomic operations are necessary since DOFs on element faces accumulate contributions from more than one element and therefore more than one thread. This does not incur any substantial serialization because only a few threads write to the same location; for the most part, threads write independently.

After the WaveHoltz iteration converges, the threads associated to boundary DOFs compute  $\lambda^{n+1}$ and write to global memory. This summarizes the one-DOF-per-thread approach.

\subsection{$T$ DOFs per Thread}\label{sec:T>1}
Now assume each subdomain has no more than $T \times K$ elements and assign to each thread block $K (P+1)^d$ threads. We assume that $T K (P+1)^d > 1024$ or that a buffer of length ${d T K (P+1)^d}$ does not fit in shared memory. For the most part, the procedure is the same for the one DOF per thread case, but each thread now stores $T$ entries of $\vec{b}, \vec{p}, \dot{\vec{p}},$ and $\vec{u}$ and the entry-wise operations in the DDM cycle are serialized by a factor $T$ compared to the one DOF per thread approach.

The key difference between the two approaches is in the computation of $A\vec{p}$. As before we have the shared memory buffers for $D$, \texttt{pShared} consisting of $K(P+1)^d$ entries, and \texttt{gradShared} consisting of $d K (P+1)^d$ entries. Note that this requires $T$ times fewer entries than above relative to the number of DOFs. Additionally, a global buffer \texttt{pGlobal} of size $L\times T K (P+1)^d$ is needed (roughly the same size as the solution vector) and each subdomain operates on a slice of length $T K (P+1)^d$ of this buffer.

\begin{figure}[!htbp]
    \centering
    \includegraphics[width=\linewidth]{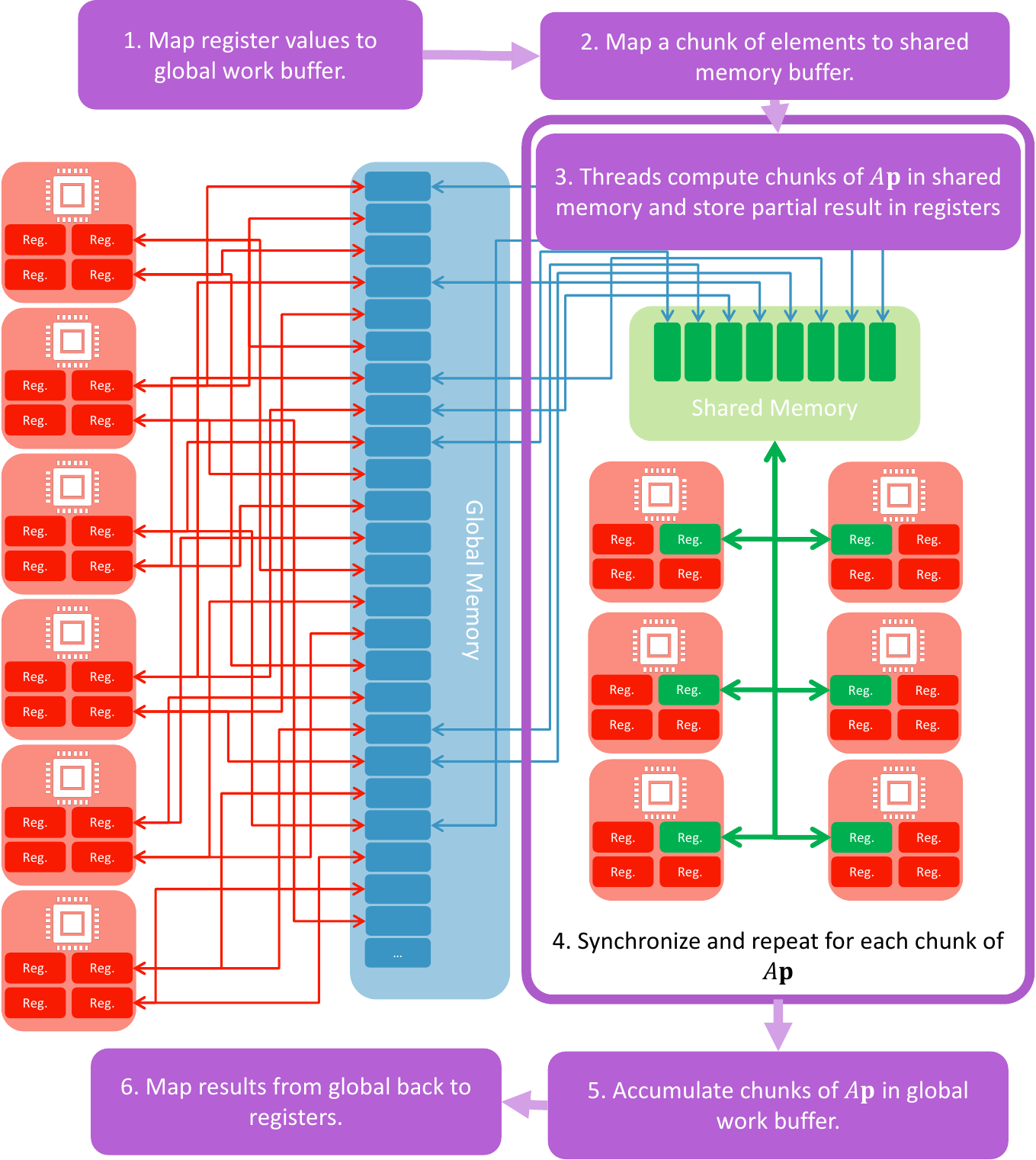}
    \caption{The $T$ DOFs per Thread algorithm with six threads and four DOFs/thread. Arrows indicate memory movement. When $T=1$ (Section \ref{sec:T=1}), the memory traffic through global is bypassed, and only the shared memory buffer is used.}
    \label{fig:tdof-alg}
\end{figure}

When evaluating $A \vec{p}$, each thread maps all $T$ values of $\vec{p}$ to \texttt{pGlobal} and the threads synchronize. The threads now work to evaluate $A \vec{p}$ in blocks of $K$ elements concurrently.
For each block of elements, the threads copy $K$ elements from \texttt{pGlobal} to \texttt{pShared}. The threads compute the entries of $A\vec{p}$ associated with this block of elements like in Section \ref{sec:T=1} but store the element contributions in registers rather than accumulate in \texttt{pShared}. The threads repeat this process $T$ times to cover all $T K$ elements resulting in $T$ entries of $A\vec{p}$ in registers. Then the threads accumulate the entries of $A\vec{p}$ in \texttt{pGlobal} atomically and synchronize. Finally, the threads collect the accumulated values of $A\vec{p}$ back from \texttt{pGlobal} into registers.
This procedure is visualized in Figure \ref{fig:tdof-alg}.

In Section \ref{sec:CUDA}, we emphasized that data movement to global memory can take as long as 300 cycles which suggests that in the worst case, the $T$ DOFs per thread strategy may be substantially more expensive than the one DOF per thread strategy. However, due to the frequent and predictable access pattern of the memory access to a fairly small slice of global memory (no more than $1024T$ values), the cache can effectively amortize the cost of the memory movement which we show in Section \ref{sec:benchmarks}.

\section{Benchmarks}\label{sec:benchmarks}
\subsection{Efficiency in Two Dimensions}
To evaluate the performance of the parallel strategies described, we measure the runtime of the DDM cycle with five WaveHoltz iterations applied to a random input vector. We compare the runtime relative to the time it takes to evaluate the Helmholtz operator $A - i \omega H - \omega^2 M$ on the whole domain applied to the same random input vector.
We consider a constant coefficient problem in $[-1,1]^2$ with uniform elements.

Since the DDM is a preconditioner, we can gain significant speedups simply by evaluating the preconditioner in a lower precision, for example as considered in \cite{lowPrecisionPrecond, tian2025mixed, guo2025adaptivemixedprecisiondynamically}. In our numerical experiments we observed roughly $1.5\times$ to $3.5\times$ speed up to be gained by evaluating the DDM cycle in single precision on an A100. Some GPU architectures such as Ada Lovelace emulate 64 bit floating point numbers and therefore achieve much lower 64 bit throughput relative to 32 bit throughput. We measured a $5\times$ to $10\times$ times speedup on an RTX 4060Ti. The following experiments are performed in single precision on an A100.

We evaluate the operators for a sequence of meshes from $32\times 32$ elements to $1024 \times 1024$ elements of polynomial degree $P \in\{ 1, 2, 3, 4, 7\}$. We vary the thread block sizes $B\in\{256, 512, 1024\}$ and the DOFs per thread $T \in\{1, 2, 4\}$. For $T = 1$ we employ the strategy discussed in \ref{sec:T=1} and for $T=2$ and $4$ we employ the strategy discussed in \ref{sec:T>1}.
For each combination of $P, B,$ and $T$ we set the number of elements per subdomain as large as possible. For example, for $P = 3, B = 1024,$ and $T = 2$ we take $\lfloor\frac{B T}{(P+1)^2}\rfloor = 128$ elements in each subdomain, e.g. as blocks of $16\times 8$ elements.

\begin{figure*}[!htbp]
    \centering
    \includegraphics[width=0.32\linewidth]{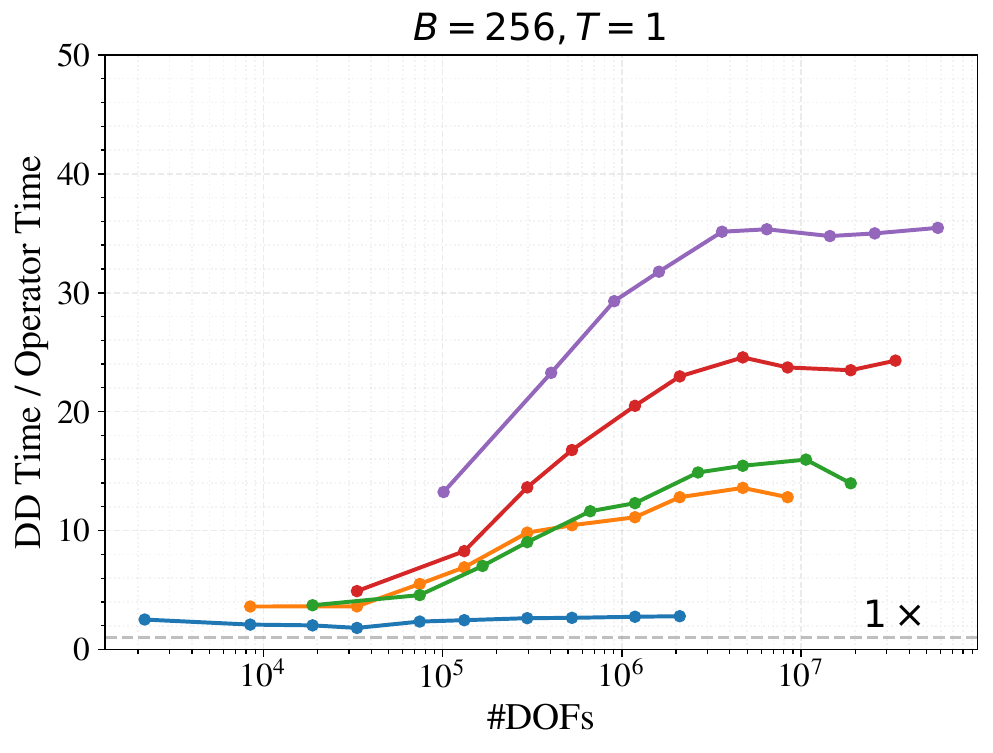}
    \includegraphics[width=0.32\linewidth]{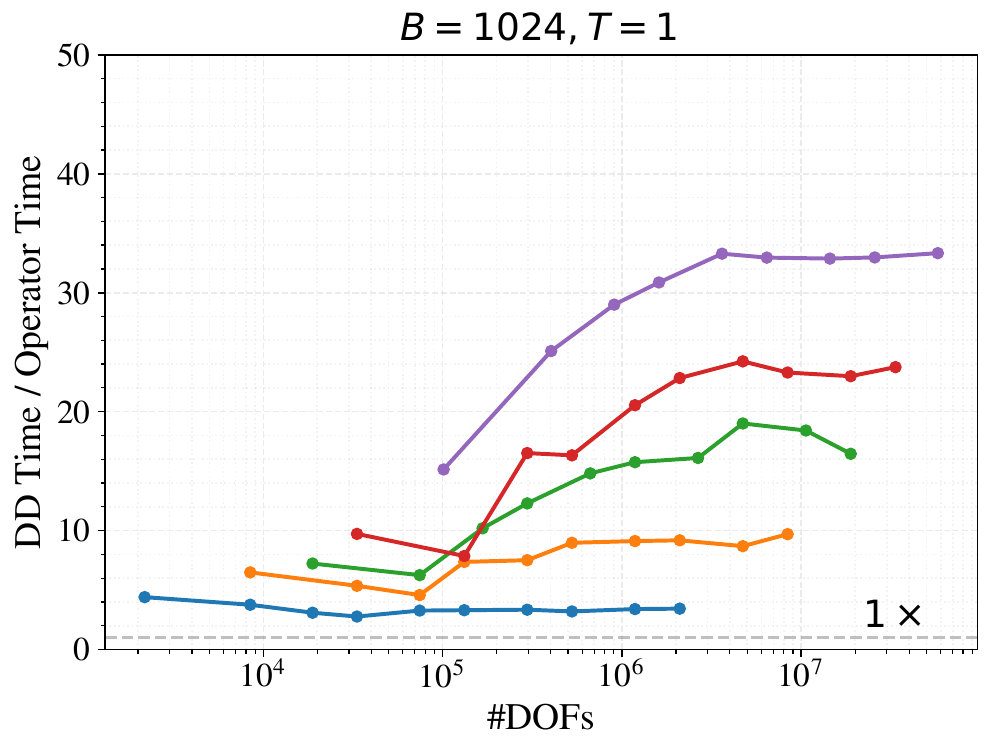}
    \includegraphics[width=0.32\linewidth]{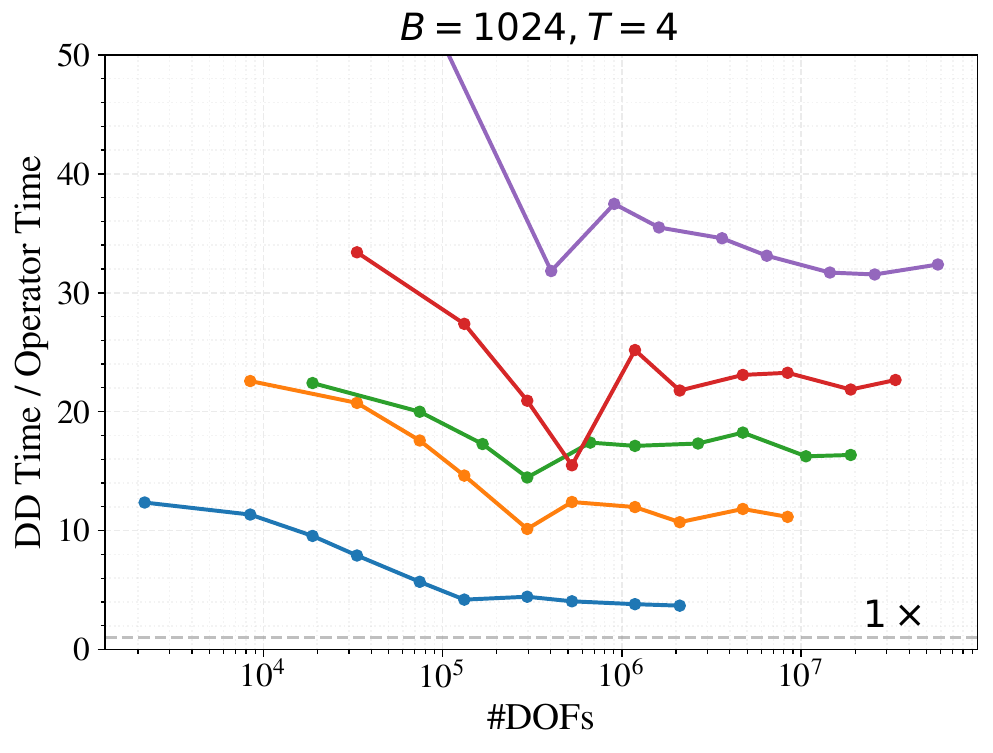}
    \includegraphics[width=0.5\linewidth]{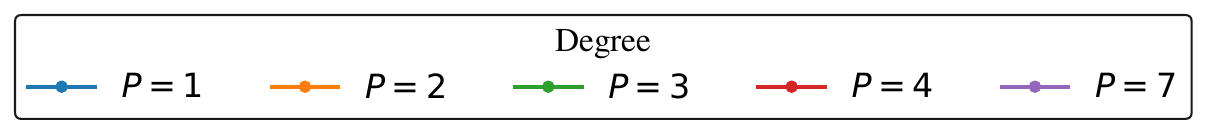}
    \caption{The time to evaluate the DDM cycle (DD time) relative to the time it takes to evaluate $A-i\omega H -\omega^2 M$ on the entire domain (Operator time) for a sequence of problems of increasing size in two dimensions.}
    \label{fig:scaling2d}
\end{figure*}

\begin{figure*}[!htbp]
    \centering
    \includegraphics[width=0.32\linewidth]{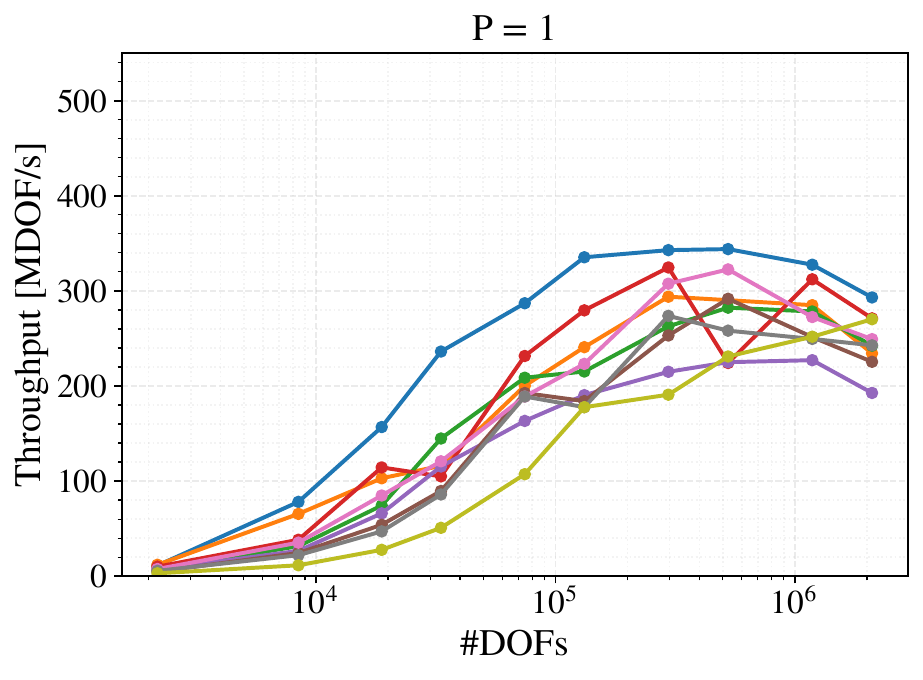}
    \includegraphics[width=0.32\linewidth]{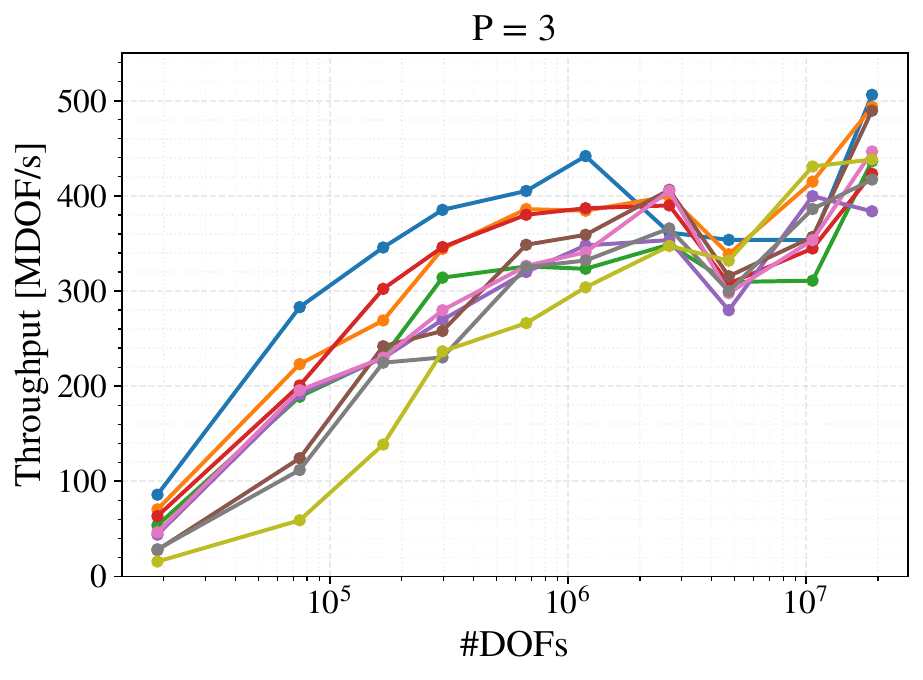}
    \includegraphics[width=0.32\linewidth]{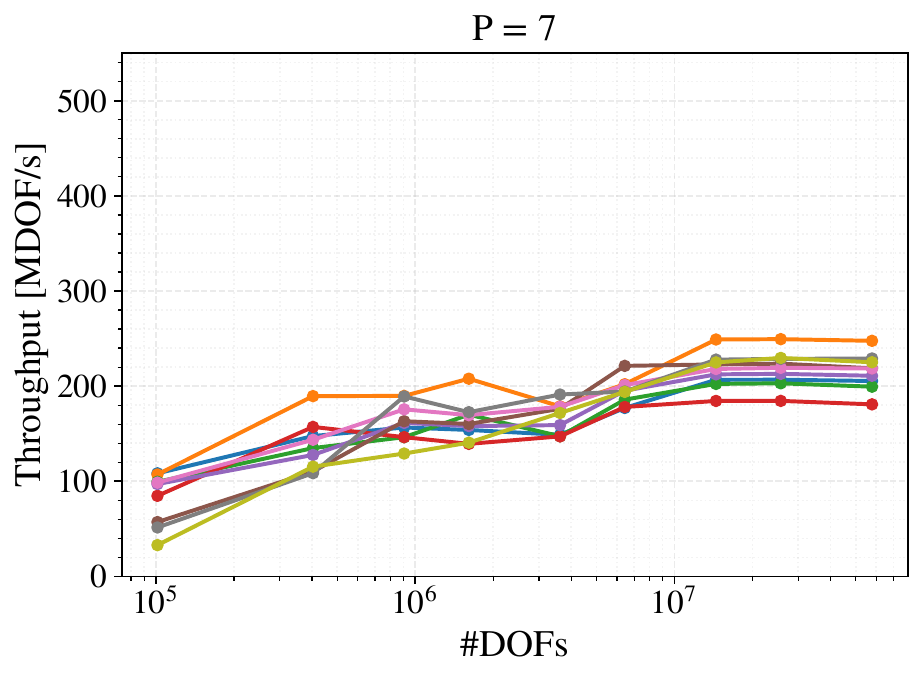}
    \includegraphics[width=0.75\linewidth]{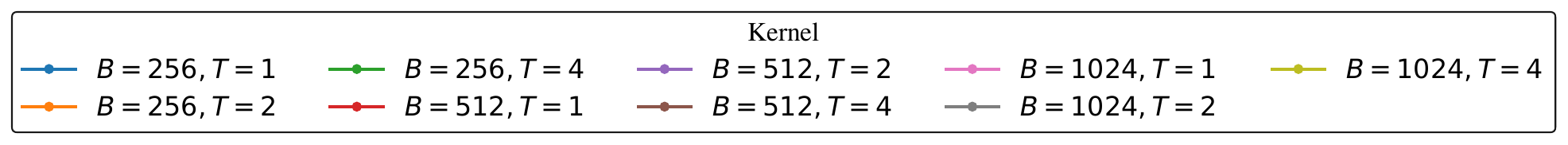}
    \caption{The throughput (millions of DOFs per second) of the various kernels for fixed degree and increasing problem size.}
    \label{fig:throughput2d}
\end{figure*}

In Figure \ref{fig:scaling2d} we plot the relative time for three kernels. Observe that the DDM is substantially more expensive for high order than lower order discretizations. We attribute this scaling to the time stepping restriction $\Delta t \lesssim 2h/p^2$. The cost of the WaveHoltz solver scales directly with $1/\Delta t$, so we expect the DDM to be more expensive for higher order discretizations\footnote{WaveHoltz has been demonstrated to be highly effective with implicit time stepping and scaling independently of polynomial degree and mesh scale (see \cite{Appel2025AnOO}) but we do not consider it here for the complexity it would introduce to the thread-block level code.}. As the problem size increases, the cost ratio tends to depend largely on degree and less so on kernel configuration. Nevertheless, kernels operating on larger subdomains appear to be less efficient for smaller problems. For example, the $B=1024,T=4$ kernel is more efficient for larger problems than smaller problem for all degrees.
Notably, the DDM cycle is very efficient for $P=1$ and can be evaluated only $2\times$ to $4\times$ the cost of the global operator.

In Figure \ref{fig:throughput2d}, we plot the throughput in millions of DOFs per second for $p=1,3,$ and $7$ for all combinations of $B$ and $T$. For $p=1$ and $p=3$ the kernels operating on the smallest subdomains are most efficient. In contrast, the most efficient kernels for $p=7$ are $B=256, T=2$ and $B=1024,T=2$ and $T=4$. It is worthwhile to note that while the smaller kernels achieve higher throughput, they serve as less optimal preconditioners and will require more iterations to converge to the solution of the global problem.

In high performance settings, particularly on GPUs, it is typically observed that higher order finite element methods are more performant due to their higher arithmetic intensity. Here however, the DDM cycle has high arithmetic intensity regardless of order, so performance is dictated by asymptotic complexity of the algorithm which is inherently higher for higher order.
Observe that the higher order $P=7$ discretization attains a peak performance around 200 MDOFs per second. The best performance is achieved by the $P=3$ discretization peaking at around 500 MDOFs per second for the range of problems considered.

\begin{figure}[!htbp]
    \centering
    \includegraphics[width=\linewidth]{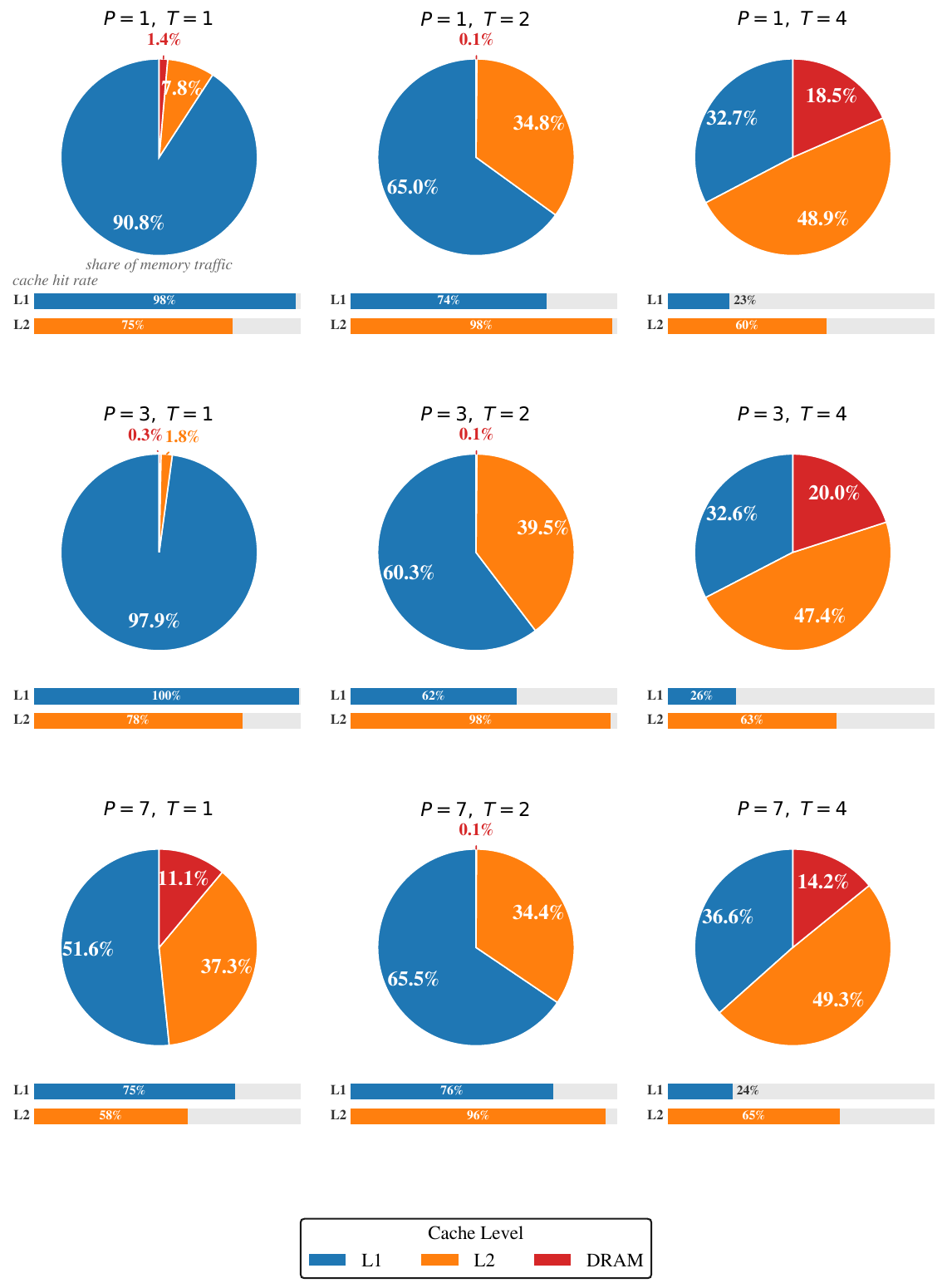}
    \caption{Share of total bytes transferred at each level of the memory hierarchy (L1, L2, DRAM) and cache hit rate for the $B=1024$ kernels for $p=1,3,$ and $7$ in two dimensions.}
    \label{fig:cache}
\end{figure}

Next, we analyze the cache usage of our method. We measure the memory traffic for the above setup for a $512\times 512$ mesh.
Figure \ref{fig:cache} shows the share of total memory movement through each level of the memory hierarchy for the $B=1024$ kernels. For $P=1$ and $3$ and $T=1$, we see that nearly all of the memory movement takes place in L1 cache with a perfect cache hit rate. For $P=7, T=1$, memory traffic spills over into L2 cache and DRAM.
For $T=4$ and all $P$, roughly one third of all data movement is through L1 and half through L2 with the rest spilling out to DRAM. This suggests that the global buffers for $T=4$ are too large to sit in cache. This is further emphasized by the cache hit rate with roughly 75\% of all L1 fetches missed. The $T=2$ kernel finds a balance. Very little memory traffic passes through DRAM, and roughly two thirds of all memory traffic is through L2 with perfect hit rate.

These results validate the design principle of the single-thread-block approach. By keeping subdomain data resident in cache, the memory overhead of the computation is substantially reduced. The $T=1$ kernel is the ideal configuration when subdomain sizes permit, and $T=2$ remains a competitive fallback for larger subdomains.

\begin{figure*}[!htbp]
    \centering
    \includegraphics[width=0.32\linewidth]{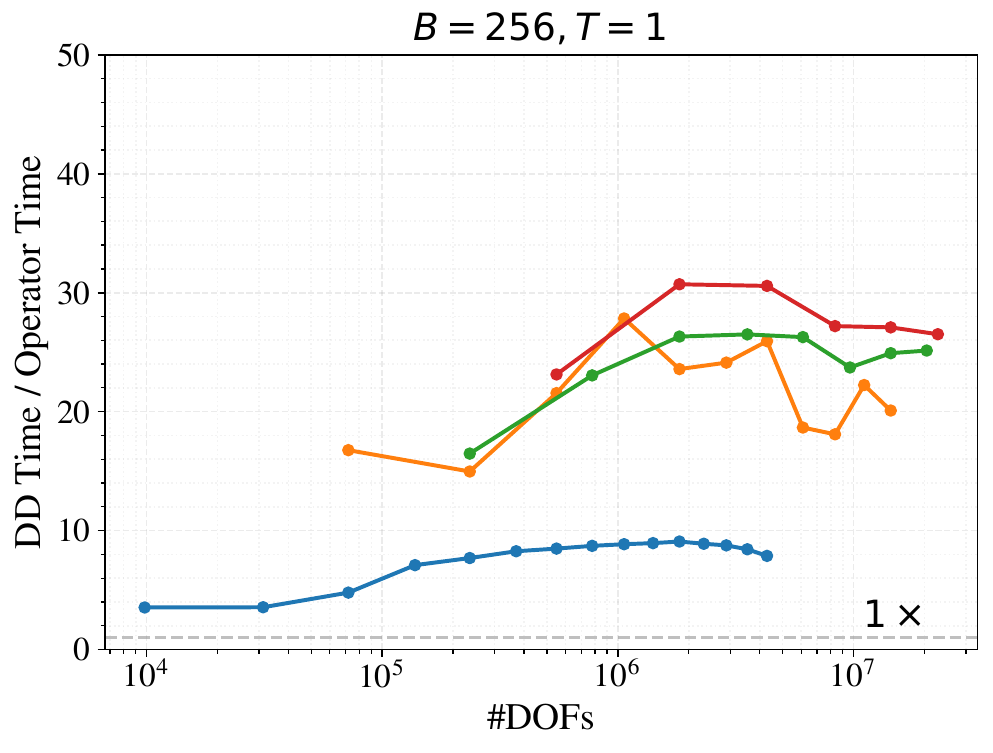}
    \includegraphics[width=0.32\linewidth]{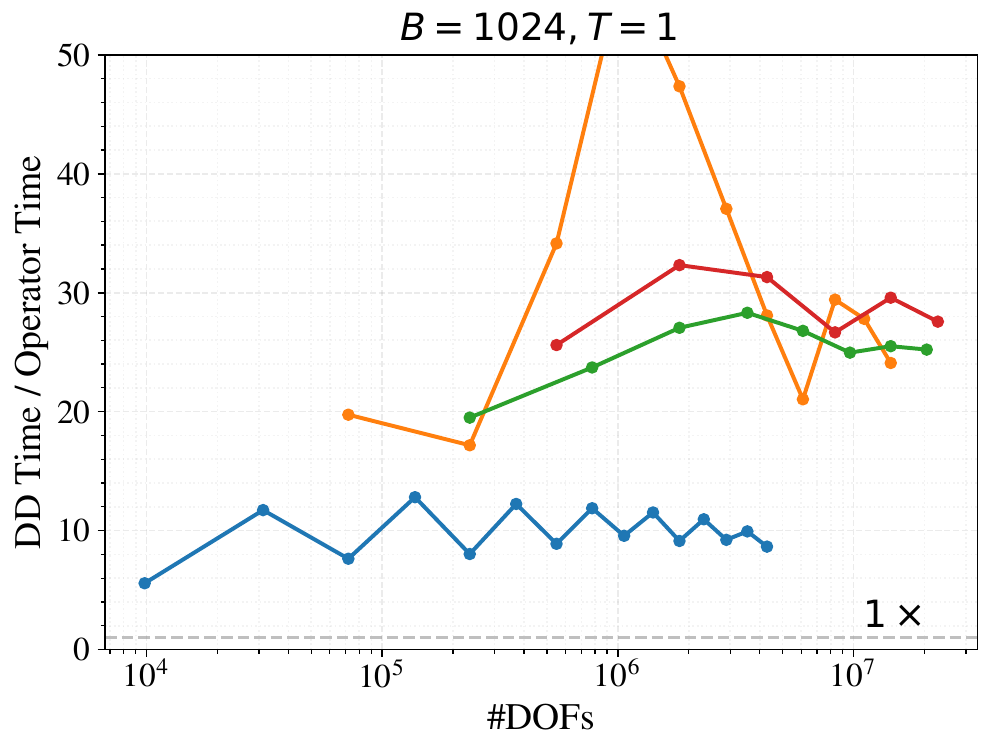}
    \includegraphics[width=0.32\linewidth]{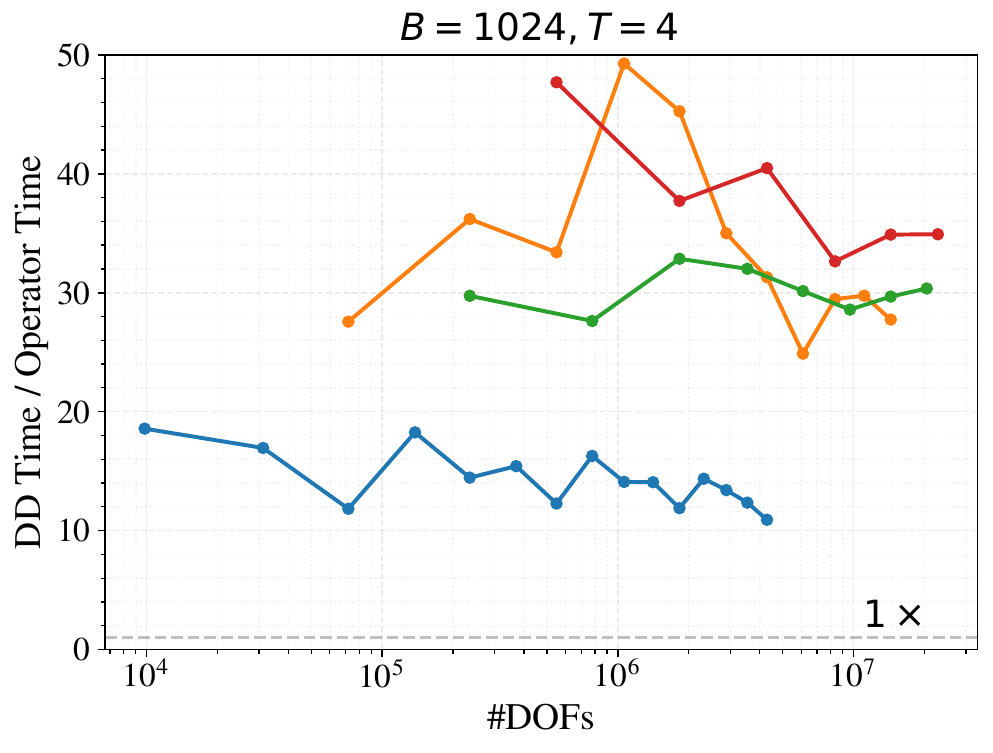}
    \includegraphics[width=0.4\linewidth]{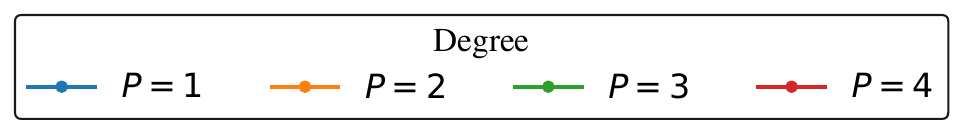}
    \caption{The time to evaluate the DDM cycle (DD time) relative to the time it takes to evaluate $A - i\omega H -\omega^2 M$ on the entire domain (Operator Time) for a sequence of problems of increasing size in three dimensions.}
    \label{fig:relative3d}
\end{figure*}

\begin{figure*}[!htbp]
    \centering
    \includegraphics[width=0.32\linewidth]{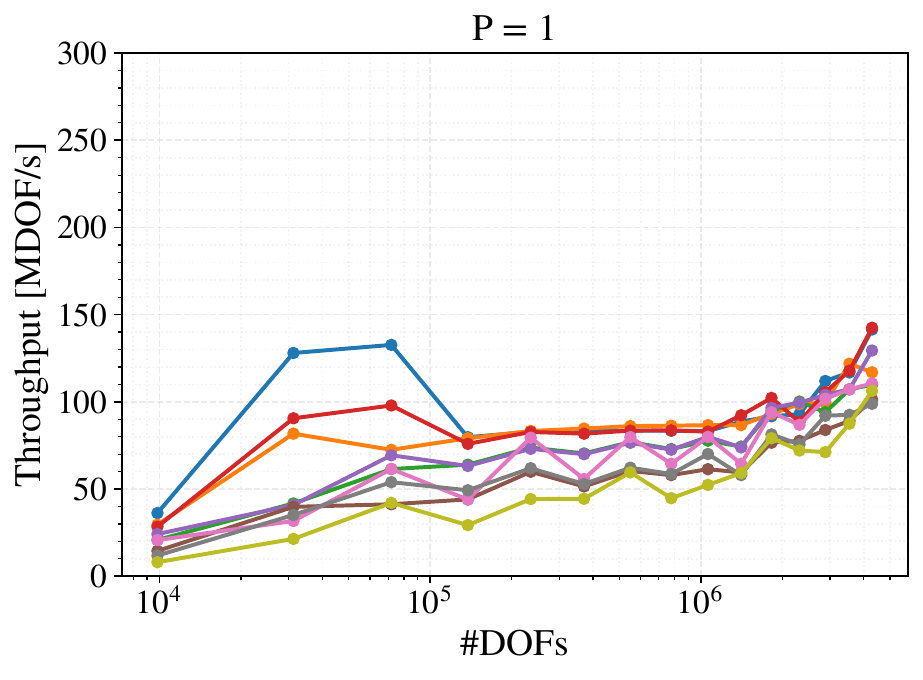}
    \includegraphics[width=0.32\linewidth]{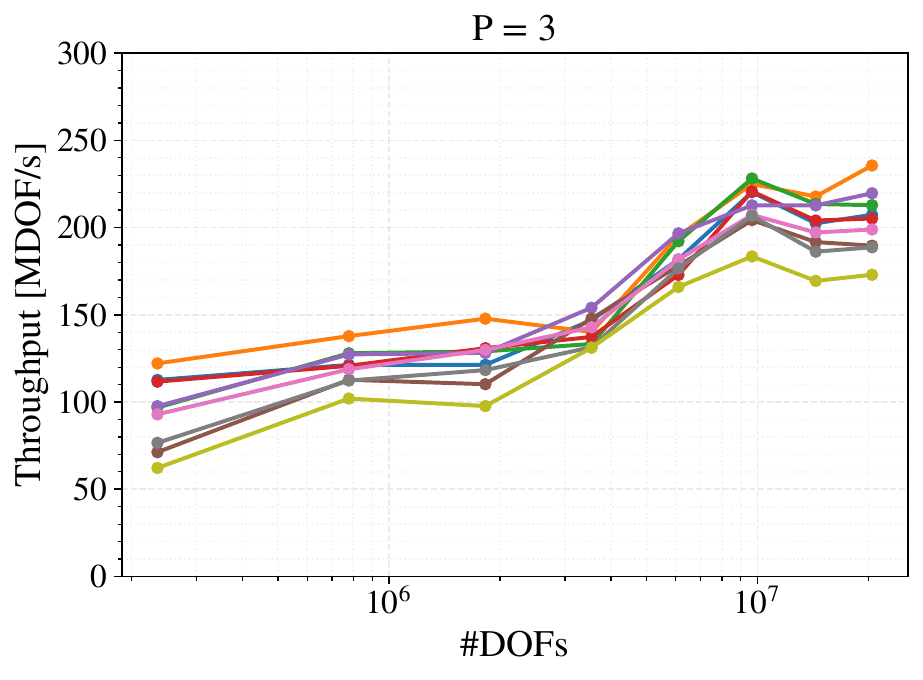}
    \includegraphics[width=0.75\linewidth]{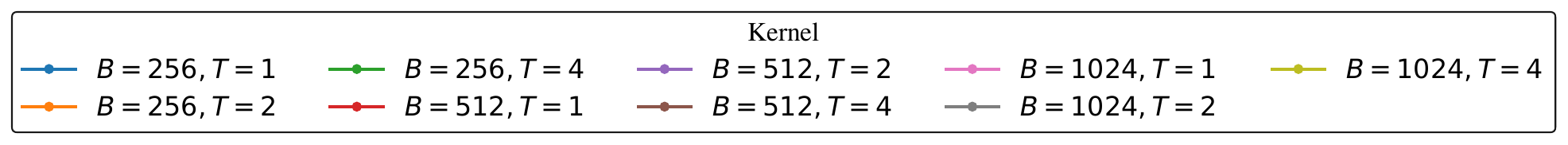}
    \caption{The throughput (millions of DOFs per second) of the various kernels for fixed degree and increasing problem size. The top panel is $p=1$ and the bottom panel is $p=3$.}
    \label{fig:throughput3d}
\end{figure*}

\subsection{Efficiency in Three Dimensions}
We repeat the two dimensional experiments experiment in three dimensions for meshes ranging from $16^3$ to $128^3$ elements. In three dimensions, the degrees of freedom per element is $(P+1)^3$ which implies that our strategy can only be applied to small subdomains. For example, $P=3$ requires 64 degrees of freedom per element, so the $B=256, T=1$ kernel configuration  can only operate on four elements per subdomain, and the $B=1024, T=4$ configuration only 64 elements. For $P=1$, we can operate on subdomains of sizes ranging from 32 elements to 512 elements. Unlike in two dimensions, we do not consider $P=7$ which requires 512 threads per element.

We compare the runtime of the DDM cycle with five WaveHoltz iterations to the time it takes to evaluate the Helmholtz operator $A-i\omega H - \omega^2 M$ on the entire domain for a constant coefficient problem in $[-1,1]^3$ with uniform elements and a random input vector. In Figure \ref{fig:relative3d}, we plot the relative time of evaluating the DDM cycle compared to evaluating the Helmholtz operator for a representative selection of $B$ and $T$ configurations. The lowest order method ($P=1$) is fastest at around $10\times$ operator evaluation time per DDM cycle. The higher order methods are more comparable to two dimensions with all three kernels ranging from $30\times$ to $40\times$ operator evaluation time for the largest problems.

In Figure \ref{fig:throughput3d}, we plot the throughput for increasing problem sizes. The method is roughly half as efficient in three dimensions compared to two.
We primarily attribute this to the DOF-per-element count scaling as $(P+1)^3$ in three dimensions, which limits the number of elements per subdomain for a fixed thread block size and reduces the arithmetic intensity of the kernel. Nonetheless, the method remains practical for low-order discretizations in three dimensions.

\subsection{Comparison with MINRES}
Next, we compare the performance of solving the subdomain problem on an A100 with WaveHoltz and MINRES. For a sequence of meshes consisting of $n\times n$ elements with $P=3$ and we take $\omega = 0.3 n$, or roughly ten elements per wavelength. Within each subdomain, both methods iterate until reaching a relative residual of $10^{-6}$ in single precision, and $10^{-12}$ in double precision.

The advantage of WaveHoltz grows with subdomain size for the following reason: as subdomains grow, the local Helmholtz problems become harder and MINRES requires substantially more iterations to converge, with each iteration requiring two inner products and five register arrays that spill into L1 cache. WaveHoltz, by contrast, requires far fewer registers and synchronizations per iteration, and is known to scale well with the problem size \cite{Appel2025AnOO}, so though each iteration is more expensive, far fewer iterations are needed. In addition, MINRES does not benefit meaningfully from mixed precision because the Krylov vectors lose orthogonality, requiring more iterations that nearly cancel out the speedup from faster floating point operations and reduced memory traffic. WaveHoltz, which essentially performs a Fourier transform in time, is numerically stable in reduced precision and benefits directly from the higher throughput.

\begin{figure}[!htbp]
    \centering
    \includegraphics[width=0.8\linewidth]{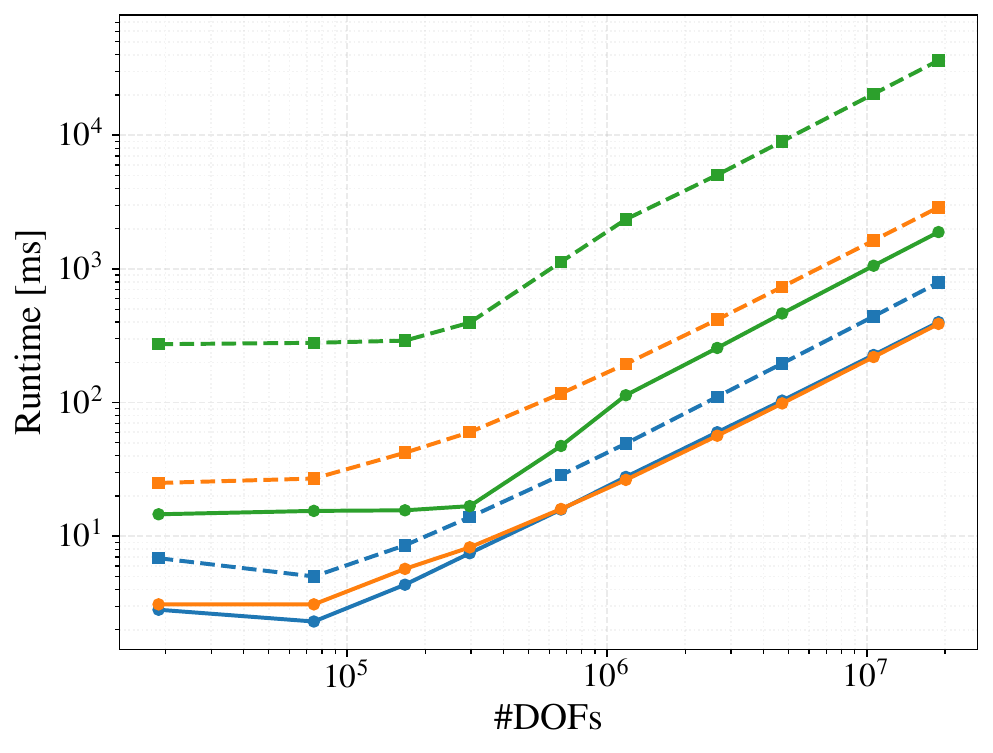} \\
    \includegraphics[width=0.8\linewidth]{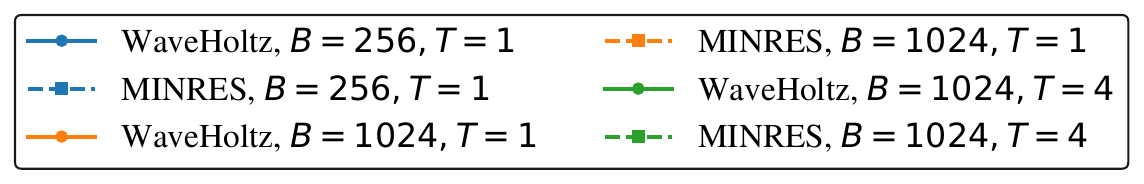}
    \caption{Runtime comparison of MINRES and WaveHoltz subdomain solvers for various kernel configurations.}
    \label{fig:minres2d}
\end{figure}

\begin{table}[!htbp]
\centering
\caption{Average speedup of the WaveHoltz subdomain solver compared to the MINRES solver for different kernel configurations in both single and double precision in two and three dimensions.}
\label{tab:speedup_results}
\begin{tabular}{cccccc}
\textbf{B} & \textbf{T} & \textbf{2D (single)} & \textbf{2D (double)} & \textbf{3D (single)} & \textbf{3D (double)} \\
\hline
\multirow{3}{*}{256} 
  & 1 & 1.97 & 1.67 & 1.58 & 0.71 \\
  & 2 & 7.25 & 2.40 & 2.85 & 2.57 \\
  & 4 & 6.52 & 2.56 & 5.03 & 5.91 \\
\hline
\multirow{3}{*}{512} 
  & 1 & 4.45 & 4.64 & 2.44 & 1.49 \\
  & 2 & 11.1 & 3.28 & 3.45 & 3.02 \\
  & 4 & 15.4 & 6.60 & 5.87 & 8.48 \\
\hline
\multirow{3}{*}{1024} 
  & 1 & 7.59 & 6.22 & 3.35 & 1.91 \\
  & 2 & 24.8 & 8.58 & 4.31 & 4.65 \\
  & 4 & 20.1 & 9.10 & 6.48 & 5.18 \\
\hline
\end{tabular}
\end{table}

In Figure \ref{fig:minres2d} we plot the runtime for three representative kernels. Across the board WaveHoltz is significantly faster, and the gap widens with subdomain size. Table \ref{tab:speedup_results} quantifies this: in single precision the average speedup ranges from $2\times$ for small subdomains ($B=256, T=1$) to $25\times$ for large ones ($B=1024, T=2$). These speedups are measured in single precision. In double precision the advantage of WaveHoltz is smaller but remains consistent: the speedup ranges from $1.6\times$ to $9\times$ and scales with the subdomain size. The experiment is repeated in three dimensions and the results are summarized in Table \ref{tab:speedup_results}. In three dimensions, the speedup ranges from $0.7\times$ to $8.5\times$ with only the ($B=256, T=1$) configuration in double precision resulting in MINRES outperforming WaveHoltz.

\section{Conclusion}\label{sec:conclusion}
We presented a procedure for accelerating non-overlapping domain decomposition for the Helmholtz equation on GPUs. Each subdomain is assigned to a single thread block, fusing the entire domain decomposition cycle into a single kernel launch. This replaces expensive global memory traffic with shared memory and register-level operations, and cache profiling confirms that memory traffic is largely confined to L1 and L2. Second, we use the WaveHoltz iteration as the subdomain solver, which requires a minimal memory footprint and minimal reduction operations, substantially reducing synchronization costs. We considered two threading strategies: one DOF per thread, suitable for subdomains up to 1024 DOFs, and $T$ DOFs per thread ($T=2,4$), which extends the approach to larger subdomains at the cost of increased cache pressure, with $T=2$ offering a good balance. Benchmarks on an NVIDIA A100 show that WaveHoltz is $2\times$ to $25\times$ faster than MINRES with the speedup growing as subdomain size increases, and that operating in single precision yields s $2\times$ to $10\times$ speedup over double precision--an advantage largely unavailable to MINRES, whose Krylov vectors quickly lose orthogonality in mixed precision.

Several extensions are natural. The zero-order transmission conditions used here can be replaced with optimized Schwarz conditions or higher order ABCs requiring minimal modification to the algorithm and is expected to substantially improve convergence of the outer iteration, particularly in three dimensions where the our method is limited to small subdomains.
Further, designing an efficient GPU algorithm for computing and applying coarse spaces by constructing planewave or eigenvector ansatz spaces on the GPU and evaluating the coarse correction within the same kernel framework, for example, is a worthwhile direction and could make the present method robust and competitive for large-scale high-frequency problems.
The variable tolerance strategy mentioned in Section \ref{sec:WH} (solving subdomain problems to a weak tolerance initially and tightening as the outer iteration approaches convergence) could substantially reduce total runtime and merits further investigation.
Finally, extension to multiple GPUs is natural: the domain decomposition structure already partitions the work into independent subdomain solves, so distributing subdomains across GPUs only requires communication of the interface data with no fundamental change to the single-GPU algorithm.

\printbibliography

\end{document}

%% file: DDAlg.tex
\KwIn{$\lambda^n$}
\KwOut{$\lambda^{n+1}$}

Assemble $\vec{b}$ from $f$ and $\lambda^n_{\ell, r}$ for each neighboring $\Omega_{r}$\;
Initialize $\vec{u}^{(0)} \gets 0$\;

\While{not converged \tcp*[r]{WaveHoltz}}{

    Initialize $\vec{p} \gets \Re\{\vec{u}^{(m)}\}$\;
    Initialize $\dot{\vec{p}} \gets \Re\{-i\omega \vec{u}^{(m)} e^{-\pi i / N_t}\}$\;

    Initialize $\vec{v} \gets \kappa_0 \vec{p} + \dfrac{\kappa_{\frac{1}{2}}}{i\omega} \dot{\vec{p}}$\;

    \For{$\nu \gets 1$ \KwTo $N_t - 1$ \tcp*[r]{time stepping}}{
        
        \ForEach{$j$}{
            $\vec{p}_j \gets \vec{p}_j + \sigma\, \dot{\vec{p}}_j$\;
        }

        Compute $A \vec{p}$ \tcp*[r]{Matrix-free}

        \ForEach{$j$}{
            $\dot{\vec{p}}_j \gets \alpha_j \dot{\vec{p}}_j 
            + \beta_j \left( -(A\vec{p})_j + \Re\{\vec{b}_j e^{-2\pi i \nu / N_t}\} \right)$\;
        }

        $\vec{v} \gets \vec{v} + \kappa_\nu \vec{p} 
        + \dfrac{\kappa_{\nu+\frac{1}{2}}}{i\omega} \dot{\vec{p}}$\;
    }

    $\vec{u}^{(m+1)} \gets \vec{v}$\;
}

$\lambda^{n+1}_{r,\ell} \gets -\lambda^n_{\ell, r} + 2 i k\, \vec{u}\big|_{\Gamma_{\ell,r}}$\;